# The same but different: impact of animal facility sanitary status on a transgenic mouse model of Alzheimer's disease

Caroline Ismeurt-Walmsley[a,1], Patrizia Giannoni[a,2], Florence Servant[b], Linda-Nora Mekki[a], Kevin Baranger[c,3], Santiago Rivera[c], Philippe Marin[a], Benjamin Lelouvier[b], Sylvie Claeysen[a]

[a] IGF, Univ. Montpellier, CNRS, INSERM, Montpellier, France
[b] VAIOMER, Labège, France
[c] Aix-Marseille Univ, CNRS, INP, Inst Neurophysiopathol, Marseille, France

[1] Present address: IGMM, Université de Montpellier, CNRS, F-34293, Montpellier, France
[2] Present address: Equipe Chrome, EA7352, Université de Nîmes, F-30000 Nîmes, France
[3] Present address: LIENSs UMRi 7266 CNRS/La Rochelle Université, F-17000 La Rochelle, France

*Correspondence to:*
Sylvie Claeysen
IGF - CNRS UMR5203 - INSERM U1191 - Univ. Montpellier
141 rue de la Cardonille, 34094 Montpellier, France
Email: sylvie.claeysen@igf.cnrs.fr

**Abstract**
The gut-brain axis has emerged as a key player in the regulation of brain function and cognitive health. Gut microbiota dysbiosis has been observed in preclinical models of Alzheimer's disease and patients. Manipulating the composition of the gut microbiota enhances or delays neuropathology and cognitive deficits in mouse models. Accordingly, the health status of the animal facility may strongly influence these outcomes. In the present study, we longitudinally analysed the faecal microbiota composition and amyloid pathology of 5XFAD mice housed in a specific opportunistic pathogen-free (SOPF) and a conventional facility. The composition of the microbiota of 5XFAD mice after aging in conventional facility showed marked differences compared to WT littermates that were not observed when the mice were bred in SOPF facility. The development of amyloid pathology was also enhanced by conventional housing. We then transplanted faecal microbiota (FMT) from both sources into wild-type (WT) mice and measured memory performance, assessed in the novel object recognition test, in transplanted animals. Mice transplanted with microbiota from conventionally bred 5XFAD mice showed impaired memory performance, whereas FMT from mice housed in SOPF facility did not induce memory deficits in transplanted mice. Finally, 18 weeks of housing SOPF-born animals in a conventional facility resulted in the reappearance of specific microbiota compositions in 5XFAD vs WT mice. In conclusion, these results show a strong impact of housing conditions on microbiota-associated phenotypes and question the relevance of breeding preclinical models in specific pathogen-free (SPF) facilities.

# 1. Introduction

Among the multiple factors involved in the development of sporadic Alzheimer's disease (AD), the identification of the microbiota-gut-brain axis provided novel insight into one of the underlying mechanisms and its potential to develop therapeutic strategies. Studies aimed at manipulating the gut microbiota of AD mouse models have shown the beneficial effect of such modulations on AD features. Indeed, antibiotic cocktails can prevent the development of amyloid burden and associated neuroinflammation in several commonly used transgenic mouse models of AD, such as APP/PS1, APP/PS1-21 and 5XFAD mice (Dodiya et al., 2019; Guilherme et al., 2021; Minter et al., 2017). Similarly, probiotic treatments prevent memory impairment in amyloid-injected murine models (Kobayashi et al., 2017; Rezaei Asl et al., 2019).

Modulating the intestinal microbiota can also be achieved by faecal microbiota transplantation (FMT), which allows engrafting microbiota into a recipient's intestinal tract. FMT has shown promising effects on AD prevention: two case reports indicated a gradual improvement in a patient's mini-mental state examination (MMSE) score assessing cognitive function, following FMT to treat a *Clostridium difficile* infection (Hazan, 2020; Park et al., 2021). Likewise, the microbiota of WT mice prevents cognitive deficits and amyloid markers in APP/PS1 and 5XFAD mice of the same age (Holsinger and Elangovan, 2020; Sun et al., 2019b). Moreover, Cryan and colleagues demonstrated that FMT from young to aged mice improves the cognitive performance of the latter (Boehme et al., 2021).

These beneficial effects prompted analyses of the microbiota composition associated with AD. It is now widely accepted that AD patients (Cattaneo et al., 2017; Guo et al., 2021; Haran et al., 2019; Li et al., 2019; Ling et al., 2020; Liu et al., 2019; Vogt et al., 2017; Zhuang et al., 2018), mouse models of amyloidosis (Chen et al., 2020c; Lee et al., 2018; Sun et al., 2019b), or of Tau pathology (Sun et al., 2019a) exhibited dysbiosis of the gut microbiota. The transfer of this dysbiotic faecal microbiota from 5XFAD mice into WT mice, induced memory impairments associated with neuroinflammation (Chen et al., 2020c; Lee et al., 2018). Altogether, this highlights a causal relationship between microbiota composition and the development of AD-associated phenotypes (Kim et al., 2021). However, the bacteria strains differ from one study to another, which underlines the importance of considering the variables affecting microbiota composition, such as animal sex, age, diet and analysis method. The sanitary status of animal housing might also have a strong influence, as specific and opportunistic pathogen free (SOPF) and specific pathogen free (SPF) housing facilities have been implemented to guarantee user safety because of the absence of numerous zoonotic pathogens, standardize animal health status, and limit variables in experiments, thus facilitating their reproducibility.

In the present study, we investigated the impact of the sanitary status of the animal facility on the faecal microbiota composition and amyloid pathology of 5XFAD mice. We also explore whether the transplantation of this dysbiotic microbiota into WT mice has an impact on cognition, depending on the breeding facility of the donor mice.

## 2. Methods

**Animal facilities**

Conventional housed mice were bred in the Faculté de Médecine Secteur Nord, INP, UMR7051 CNRS/AMU, Marseille, France, and SOPF mice in the Plateau Central d'Élevage et d'Archivage, CNRS, Montpellier, France. Micro-organisms excluded in this SOPF animal facility are listed in Supplemental Table 1. WT recipient mice, *i.e.*, C57Bl/6J were purchased from Janvier-Labs (France).

**Mice**

5XFAD mice (Ismeurt et al., 2020; Oakley et al., 2006) overexpress human amyloid precursor protein gene (*APP*) bearing Swedish (K670N, M671L), Florida (I716V), and London (V717I) familial AD mutations, and, human presenilin 1 gene (*PS1*) carrying M146L and L286V mutations, both under the control of the neuronal mouse Thy1 promoter. The 5XFAD strain (B6/SJL background) was backcrossed more than 10 generations in C57BL/6J mice by crossing hemizygous transgenic mice with C57BL/6J F1 breeders (Janvier-Labs, Le-Genest-Saint-Isle, France). Litters used in the current study were obtained by crossing hemizygous 5XFAD male with WT females, thus ensuring that pups were born with a naive microbiota. Mice were housed in the respective facilities for more than 3 generations. At weaning (P21), 5XFAD$^{Tg}$/$_0$ and WT littermates were separately caged (three to five animals) to avoid coprophagy between animals of different genotypes. Mice were kept with a 12 h day/night cycle (120-150 LUX) at 21-24°C and 50 ± 10 % humidity. Food and water were available *ad libitum* (SAFE® A03, A04 SAFE, Augy, France). SOPF water was softened to TH 3.5/4°fH, UV treated and filtered to 0.2µM. Tap water was used in conventional housing. In SOPF facility, mice were housed in ventilated 501 cm² cages (GM500, Techniplast, Decines Charpieu, France) on poplar bedding (SAFE® select fine, SAFE, Augy, France). In conventional facility, mice were housed on similar bedding in 530 cm² cages without filtering lid on standard racks, (1284L EUROSTANDARD TYPE II L, Techniplast, Decines Charpieu, France). All animal care and experimental procedures were performed in accordance with National and European regulations (EU directive N°2010/63) and the care guidelines of Montpellier University (authorization B34-417-24, approved protocol #21222).

**Faecal microbiota preparation and administration**

The faecal microbiota was sampled at 2 months of age, when amyloid accumulation began but before the onset of cognitive deficits in 5XFAD mice, and at 6 months of age, when the amyloid load was intense and cognitive deficits had set in (Oakley et al., 2006). Following sacrifice of the donor mice (n = 4-8, gender-mixed), the faecal contents of the distal and proximal colon were processed within 2 h of sampling. The material from all animals was pooled, weighed and diluted in sterile PBS at 100 mg/mL. Samples were homogenised in a vortex mixer for 3 min, centrifuged at 800 × $g$ for 2 min at 4°C, aliquoted, and stored at -20°C. Each day of treatment, aliquoted faecal material was diluted 1:10 with sterile PBS, then WT female C57Bl/6J mice received 100 µL by oral gavage. A sham group of control mice received 100 µL of sterile PBS. Mice were treated twice a week from 8 to 16 weeks of age.

**Novel object recognition test**

The novel object recognition (NOR) test (Bevins and Besheer, 2006) was carried out in an open-field in a dimly lit room (20-30 Lux). On day 1, mice were habituated to the empty arena for 10 min. On day 2, mice were trained to explore two objects (familiar objects) for 10 min and then returned to their home cage. On day 3, they were subjected to a 5-min restitution session in which one of the two (familiar) objects was replaced by a new one, *i.e.,* novel object. All sessions were video recorded, and analyses were blindly performed to quantify the time the mouse actively explored each object. The discrimination index corresponding to [(exploration time of novel object – exploration time of familiar object)/total exploration time] was then calculated.

**Amyloid load analysis**

Thirty-five-µm coronal sections of mice brains were cut on a cryostat, then mounted on slides (SuperFrost Ultra Plus®, Epredia, France) in Tris-Gelatin and dried at room temperature (RT) for 24 h. The following day, slides were washed in deionized $H_2O$ ($H_2Od$), dehydrated, then incubated in a 1:1 mixture of 96% EtOH (CarloErba, France) and Chloroform (HoneyWell, Germany) for 30 min at RT. The slides were then rehydrated, incubated in 0.1% Thioflavin S (Sigma-Aldrich, France) for 10 min, washed in 70% EtOH for 5 min, and then in $H_2Od$. The nuclei were stained with DAPI before the slides were coverslipped (Marienfeld, Germany) with mounting medium (Pertex, Histolab, Sweden). Images were acquired with an upright fluorescence microscope (Leica Thunder, sCMOS Leica DFC9000 camera) with a 10x objective lens. For each section, the whole hippocampus and parietal cortex were blindly examined to determine the number and the total area of amyloid deposits for each region. Two-three sections per animal were analysed.

**16S rRNA targeted metagenomic analysis**

Individual faeces were collected from donor mice, frozen in liquid nitrogen and stored at -80°C until use. The bacterial population present in the samples was determined by targeted sequencing of the V3-V4 variable regions of the 16S rRNA gene according to a protocol established by Vaiomer, France. The workflow classification is restricted to bacteria. The extraction steps, library construction, sequencing and bioinformatics pipelines have been previously described (Boulanger et al., 2023; Lluch et al., 2015). The targeted microbiome sequences were analysed using the bioinformatics pipeline (including *LEfSe* - Linear discriminant analysis Effect Size) established by Vaiomer based on the Find Rapidly OTUs with Galaxy Solution (FROGS) guidelines (Boulanger et al., 2023; Escudie et al., 2018).

**Statistical analysis**

The impact of FMT on object time exploration were determined by ANOVA followed by post-hoc test (Bonferroni's), after verification of Gaussian distribution (Shapiro–Wilk normality test), and homogeneity of sample variance (Brown-Forsythe's and Bartlett's tests). For discrimination indexes, Student's t-test was used. For all statistical tests, a $p < 0.05$ was considered significant. Analyses were performed using Prism 9 (GraphPad, MA, USA). For metagenomic analyses, Mann-Whitney test was used when comparing two groups, and Kruskal-Wallis test for comparisons involving more than two groups.

## 3. Results

**Dysbiotic faecal microbiota in 6-month-old, conventionally bred 5XFAD mice**

To investigate the impact of sanitary status on the composition of the microbiota of 5XFAD mice, we first conducted 16S rRNA targeted metagenomic analysis of faecal samples collected from 2- and 6-month-old mice breaded in conventional housing. At 2 months of age, 5XFAD mice showed a trend towards decreased alpha diversity for all indexes, compared to WT littermates (Fig. 1A and Supplementary Fig. 1A). The Chao1 index reflects the richness of the samples, while the Shannon and Simpson indexes reflect both the richness and evenness of the diversity of samples (Shannon being more sensitive to evenness and Simpson to richness). Beta diversity of each group, illustrated by Principal Coordinate Analysis (PCoA) of the Bray-Curtis index (comparing dissimilarity between samples based on species abundance), showed no significant difference (PERMANOVA & PERMDISP) between the bacterial composition of 5XFAD and WT mice (Fig. 1B and Supplementary Fig. 1A). Statistical pairwise comparison using LEfSe analysis revealed only a few significantly different bacterial phylotypes compared to WT littermates (Fig. 1C), as did the relative abundance of bacterial genera (Fig. 1D).

At 6 months of age, 5XFAD mice showed a significant increase in alpha diversity for the Chao1 (p = 0.04), Shannon (p = 0.02) and Simpson (p = 0.02) indexes (Fig. 1A and Supplementary Fig. 1B), compared to WT mice. Beta diversity revealed significant dissimilarity of bacterial populations in the two genotypes in both Bray-Curtis (Fig. 1B, PERMANOVA p = 0.02) and Unifrac indexes (Supplementary Fig. 1B, PERMANOVA p = 0.03, PERMDISP p = 0.03). Statistical pairwise comparison and relative abundance of bacterial genera (Fig. 1C & D) confirmed that 6-month-old 5XFAD mice had a marked different bacterial composition, compared to WT littermates of the same age, presenting a global reduced proportion of Bacteroidota phylum which has been documented by others (He et al., 2024).

Globally, the relative abundance of the top 20 genera within the annotated bacteria (Fig. 1D) showed that an unknown genus belonging to the *Muribaculaceae* family was the most abundant (40% of top 20 genera in WT and 5XFAD 2-month-old mice; increasing to 62% in WT 6-month-old mice; and decreasing to 23% in 5XFAD 6-month-old mice). We also found 2-fold and 31-fold increase in *Prevotellaceae UCG-001*, the second most abundant genus, in 5XFAD mice compared to WT mice at 2 and 6 months, respectively. Of note, we also detected in all groups the clear presence of *Helicobacter* genus (0.8% to 8%), a bacterial genus that is excluded from animal facilities with restricted sanitary status (Supplementary Table 1). In other studies, significant decreases in *Muribaculum* genus or *Muribaculaceae* family were similarly observed in 6–8-month-old 5XFAD mice, which also showed an increase in the *Helicobacter* genus, suggesting that they were probably housed in conventional facilities (He et al., 2024; Luo et al., 2022).

Taken together, these metagenomic analyses showed that 5XFAD mice in conventional housing develop a dysbiotic faecal microbiota with age, compared to their WT littermates.

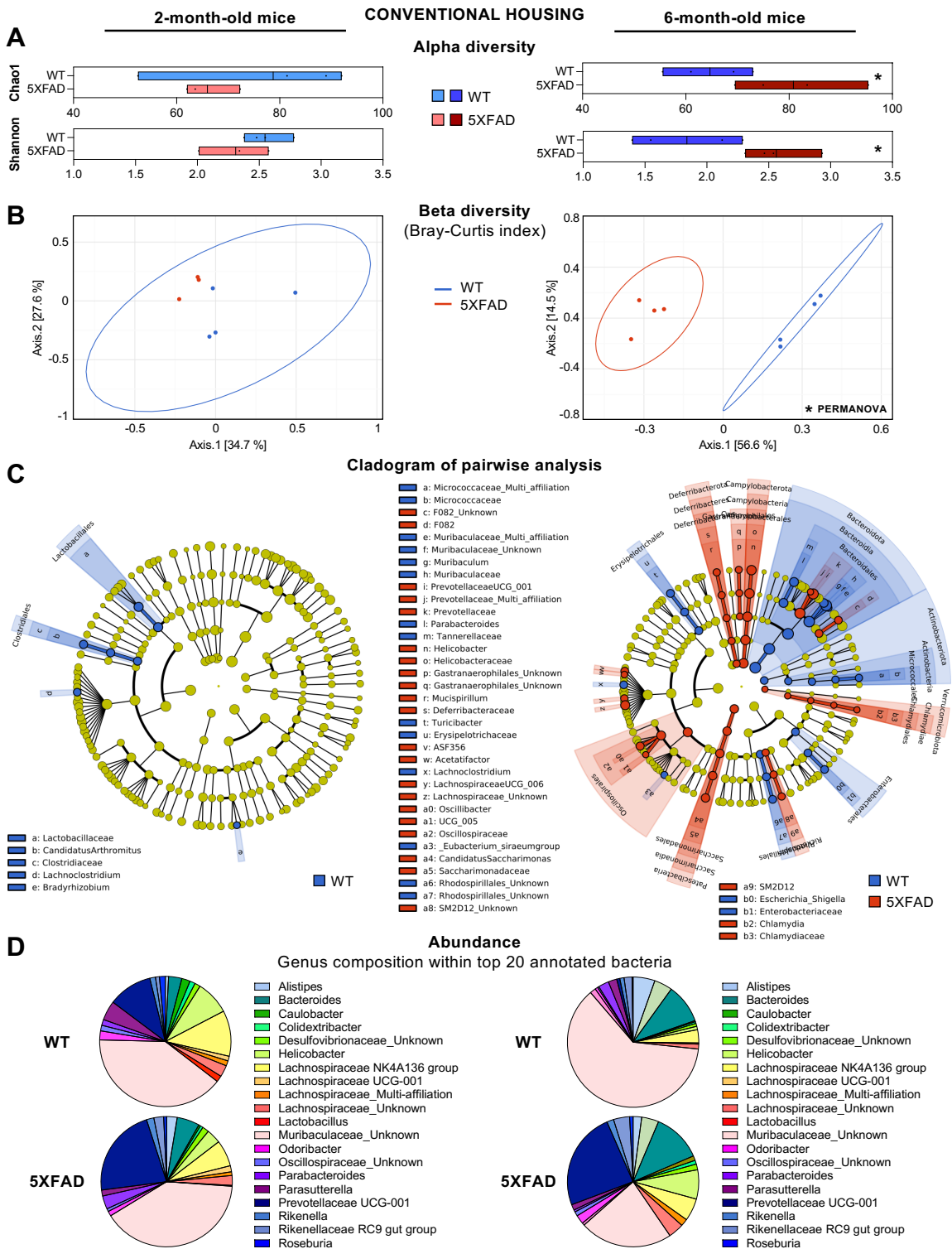

**Figure 1. Faecal microbiota composition of 5XFAD and WT littermates in conventional housing.**

**(A)** Alpha diversity indexes at the genus level. Data are presented as minimum to maximum and mean. Each dot represents a mouse (n = 3-4). *p < 0.05 versus WT (Kruskal-Wallis) **(B)** Beta diversity assessed by principal coordinates analysis using the Bray-Curtis index. *p < 0.05 versus WT (PERMANOVA). **(C)** Cladogram showing the significant increase in OTU and bacterial taxa identified by LEfSe analysis. In blue: more abundant in WT; in red: more abundant in 5XFAD. **(D)** Synthetic diagram presenting the relative abundance of top 20 annotated genera. Left panels: data from 2-month-old mice. Right panels: data from 6-month-old mice.

**Absence of dysbiosis of faecal microbiota in 6-month-old, SOPF-bred 5XFAD mice**

We then carried out 16S rRNA targeted metagenomic analysis of faecal samples from 5XFAD mice bred in SOPF facility to monitor the development of dysbiosis between the ages of 2 and 6 months. At both ages, we found no significant difference in either alpha diversity (using Chao1, Shannon and Simpson indexes) or beta diversity (using the Bray-Curtis and Unifrac indexes) between WT and 5XFAD mice (Fig. 2A-B and Supplementary Fig. 1B). At 6 months of age, the bacterial composition was still highly similar in both groups (Fig. 2B). Statistical pairwise comparison and relative abundance of taxa at genus level confirmed no major divergence between the two phenotypes (Fig. 2C-D). In the SOPF facility, an unknown genus belonging to the *Muribaculaceae* family was still the most abundant among the top 20 annotated genera (Fig. 2D). We observed no change in the relative abundance of *Prevotellaceae UCG-001* genus in 6-month-old 5XFAD mice compared to WT mice (3% *vs* 5%).

Together, these analyses indicate that a SOPF environment prevents the development of a dysbiotic faecal microbiota in 6-month-old 5XFAD mice, compared to WT littermates.

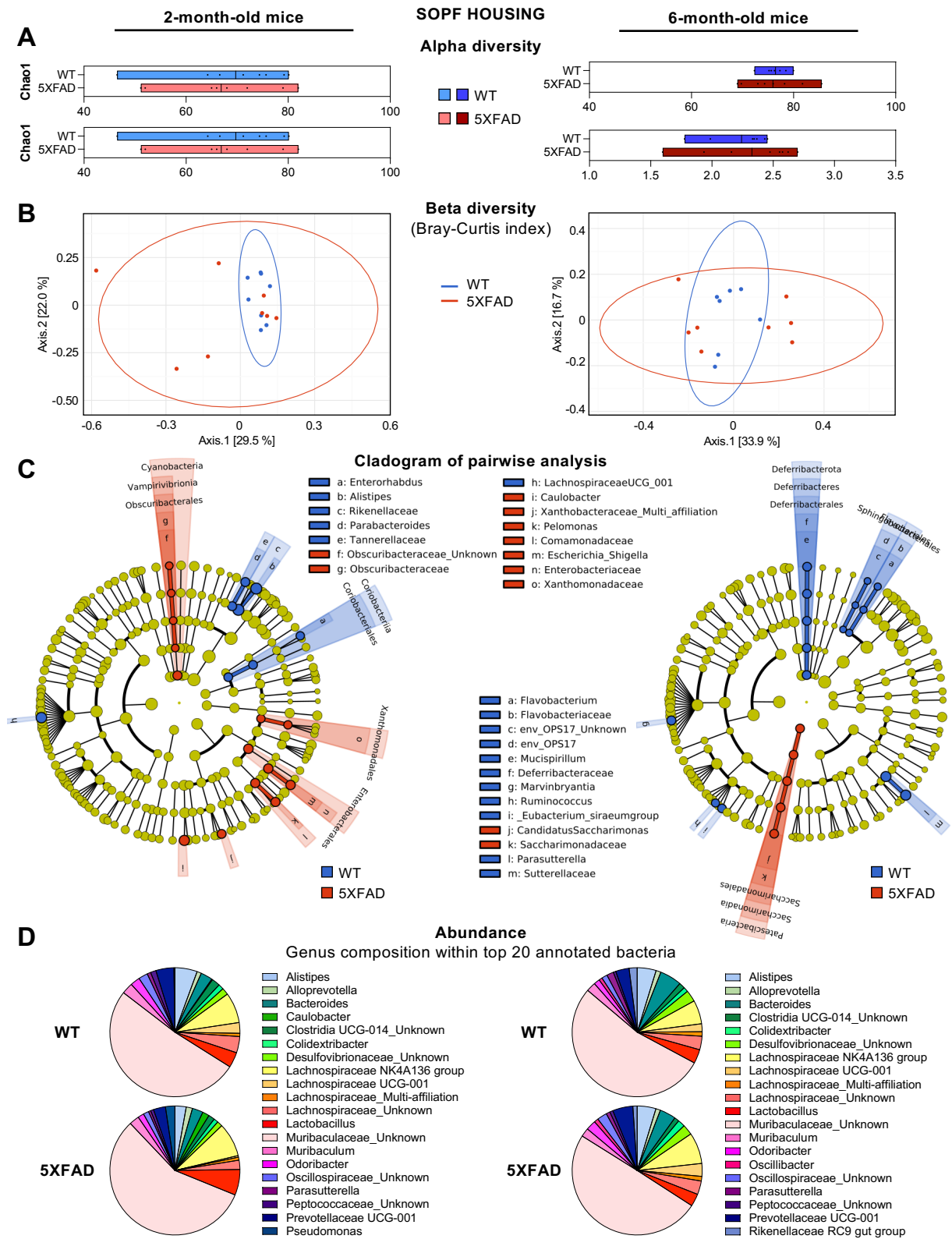

**Figure 2. Faecal microbiota composition of 5XFAD and WT littermates in SOPF housing.**

**(A)** Alpha diversity indexes at the genus level. Data are presented as minimum to maximum and mean. Each dot represents a mouse (n = 7-8). **(B)** Beta diversity assessed by principal coordinates analysis using the Bray-Curtis index. **(C)** Cladogram showing the significant increase in OTU and bacterial taxa identified by LEfSe analysis. In blue: more abundant in WT; in red: more abundant in 5XFAD. **(D)** Synthetic diagram presenting relative abundance of the top 20 annotated genera. SOPF: Specific and Opportunistic Pathogen-Free. Left panels: data from 2-month-old mice. Right panels: data from 6-month-old mice.

**Impact of donor housing on memory alteration by chronic transplantation of faecal microbiota**

To determine whether these distinct microbial compositions have a differential impact on cognition following FMT to WT mice, faecal microbiota from 2- or 6-month-old 5XFAD mice bred in either a conventional or SOPF environment, were administered to WT mice by chronic gavage for 2 months (Fig. 3A). The cognitive performance of the recipient mice was assessed at 16 weeks of age using the NOR test. Of note, experiments with conventionally reared animals at 2 and 6 months of age were performed together, thus having the same control group (vehicle treated mice), whereas experiments with SOPF breaded animals were performed in 2 batches, thus having independent control groups.

WT recipient mice that received vehicle solution presented normal memory performance, as shown by the higher percentage of exploration time spent exploring the novel object (Fig. 3B, D) and the positive discrimination indexes (Fig. 3C, E). The transplantation of faecal microbiota from conventionally housed 5XFAD mice aged 2 or 6 months induced cognitive deficits in the recipient mice (Fig. 3B, C). In contrast, mice receiving microbiota from 2- or 6-month-old 5XFAD mice bread in SOPF facility showed similar memory performance to that of mice transplanted with vehicle, as shown by the significant difference in percentage of exploration time and discrimination indexes (Fig. 3D, E). Collectively, these results demonstrate that the faecal microbiota of 5XFAD mice differentially impacts the cognition of WT recipient mice in the NOR test, depending on the initial housing in conventional *vs* SOPF facility.

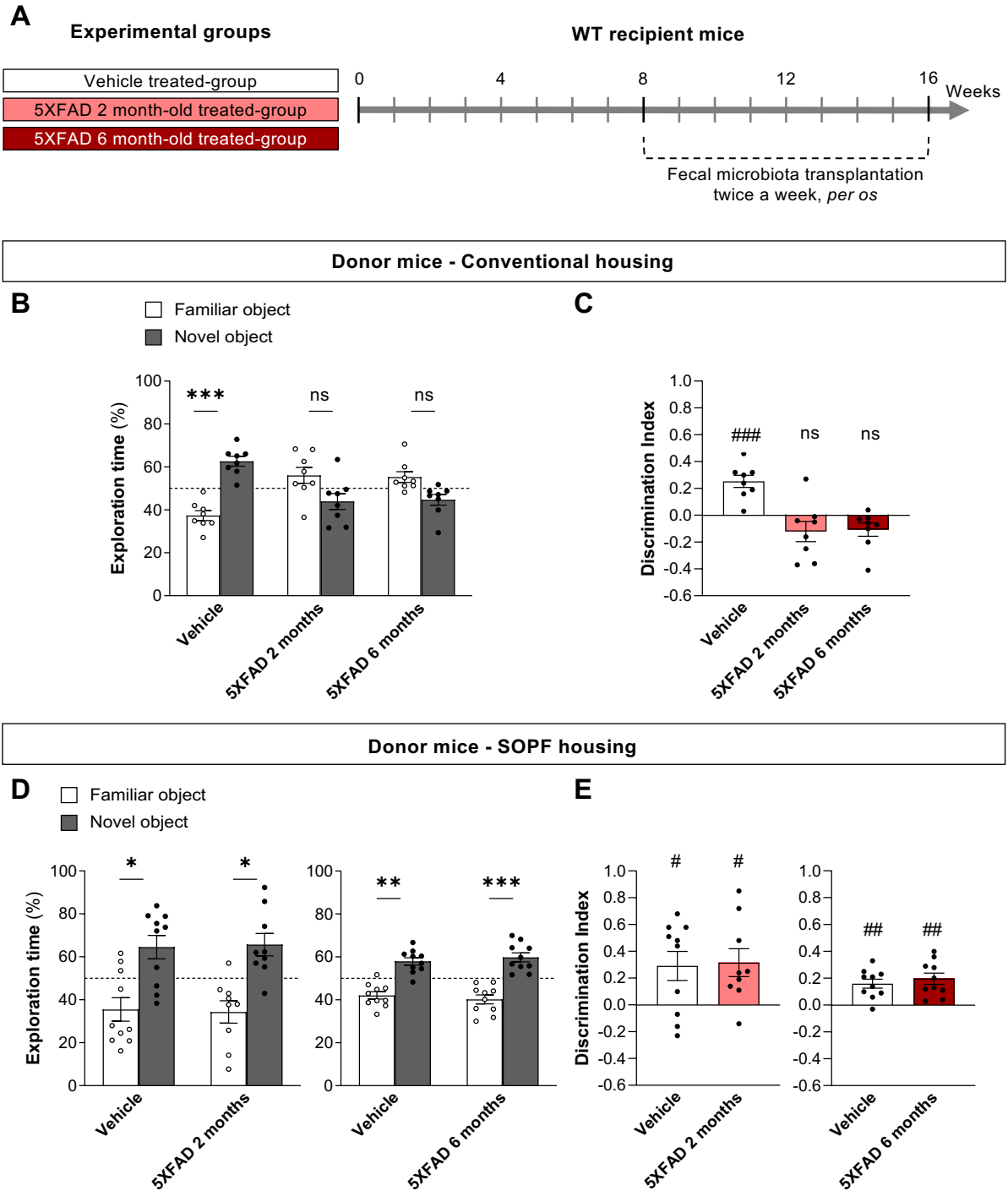

**Figure 3. Impact of donor housing conditions on cognitive performance alteration induced by FMT**

**(A)** Experimental design. **(B-E)** Novel object recognition test in 16-week-old mice performed with a 24-h intersession interval. **(B, D)** Time spent by mice exploring familiar and novel objects during test session. ***$p < 0.001$, **$p < 0.01$, *$p < 0.05$, significantly different from familiar object (Two-way repeated-measures ANOVA, followed by Bonferroni's test). **(C, E)** Discrimination index calculated by using exploration times in the test session and formula [(novel − familiar) / (familiar + novel)]. Indices positively different from zero: ##$p < 0.01$, #$p < 0.05$ (One sample Student's t test). Data are presented as means ± SEM. Each dot represents a mouse (n = 8-10). As a result of two distinct experiments, the vehicle groups are presented separately in D and E panels. SOPF: Specific and Opportunistic Pathogen-Free.

**Re-establishment of a dysbiotic microbiome in SOPF-born 5XFAD mice after 18 weeks in a conventional environment**

As co-housing SPF-born mice with conventionally-born mice in a conventional facility significantly increased their gut microbiota diversity (Chen et al., 2020b), we investigated whether SOPF-born 5XFAD mice develop a dysbiosis of faecal microbiota compared to WT littermates after 18 weeks of housing in a conventional environment. Bacterial composition of faecal samples showed no significant modification in alpha diversity for 5XFAD mice after 18 weeks in a conventional environment, compared to WT mice (Fig. 4A), whereas Bray-Curtis dissimilarity and Unifrac distances revealed significant differences in the two groups (PERMDISP, $p = 0.0099$ and $p = 0.02$; PERMANOVA, $p = 0.1$ and $p = 0.02$, respectively, Fig. 4B and Supplementary Fig. 2). The relative abundance at the genus level and pairwise statistical comparisons showed that WT and 5XFAD mice are differentially colonised by bacteria living in conventional environment (Fig. 4C, D) and that the composition of their respective microbiota shows greater relative differences than in the SOPF housing (Fig. 2C, D). Taken together, these results demonstrate that rearing SOPF-born animals in conventional housing allows the development of a dysbiotic faecal microbiota in 6-month-old 5XFAD mice.

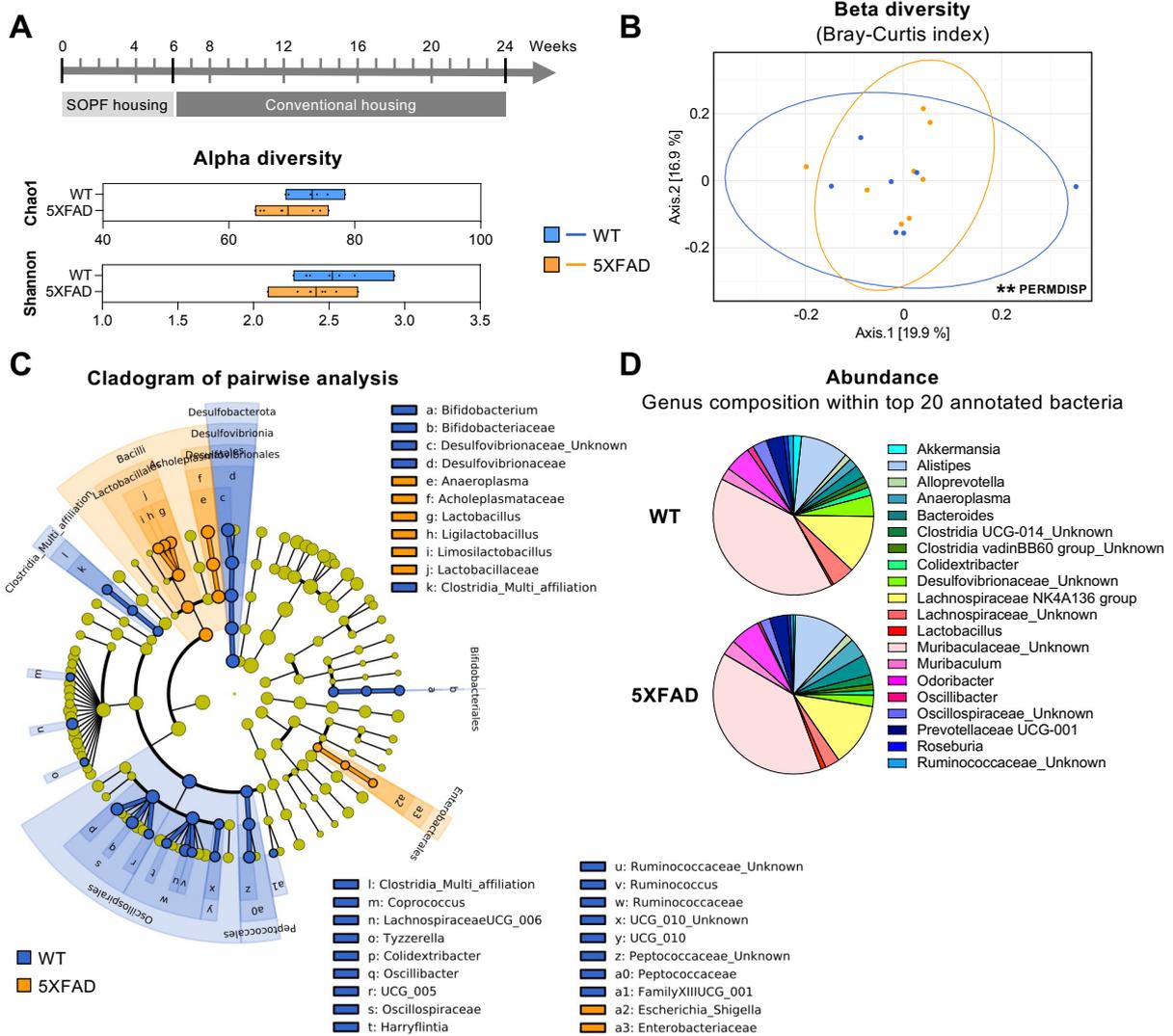

**Figure 4. Faecal microbiota composition of SOPF-born 5XFAD and WT littermates after 18 weeks in conventional housing.**

**(A)** Alpha diversity indexes at the genus level. Data are presented as minimum to maximum, and mean. Each dot represents a mouse (n = 7-8). **(B)** Beta diversity assessed by principal coordinates analysis using the Bray-Curtis index. *p < 0.05 versus WT (PERMDISP). **(C)** Cladogram showing the significant increase in OTU and bacterial taxa identified by LEfSe analysis. In blue: more abundant in WT; in orange: more abundant in 5XFAD. **(D)** Synthetic diagram showing the relative abundance of the top 20 annotated genera.

## Reduction of amyloid pathology in 6-month-old SOPF-bred 5XFAD mice compared to conventionally bred 5XFAD mice

Since germ-free AD mouse models are known to produce less Aβ peptides and deposits in the brain compared to conventionally bred mice (Harach et al., 2017), we investigated whether 6 month-old 5XFAD mice bred in the SOPF housing exhibit lower brain amyloid loads compared to 5XFAD mice of the same age bred in the conventional facility. Histopathologic staining of compact amyloid deposits with Thioflavin S was performed on hippocampal and cortical sections (Fig. 5A-E). The number of amyloid deposits was significantly reduced in the cortex of SOPF-5XFAD mice compared to conventional 5XFAD mice (-19%, p = 0.025, Fig. 5C). Similarly, a significant decrease in the total area of cortical thioflavin staining was observed in the brains of SOPF-5XFAD mice (-18%, p = 0.033, Fig. 5E). In the hippocampus, no significant effect was observed for any parameter, although half of the animals showed a downward trend (Fig. 5B, D).

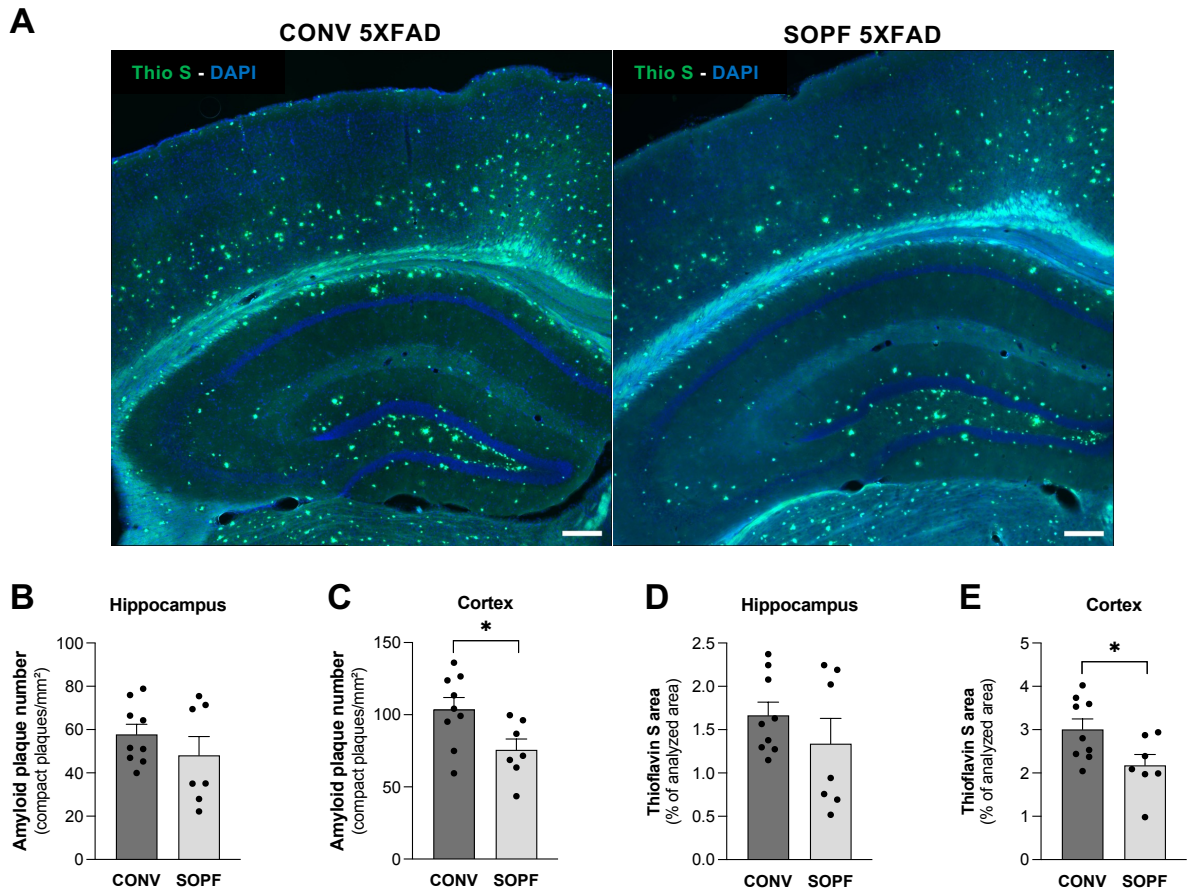

**Figure 5. Amyloid load in 6-month-old 5XFAD mice bred conventionally and in SOPF housing.**

**(A)** Representative images of compact amyloid plaques stained with Thioflavin S (green) and nuclei stained with DAPI (blue). Scale bar: 200 µm, magnification 10x. **(B-E)** Quantification of the number **(B, C)** and area **(D, E)** of amyloid aggregates normalized to the total area of the hippocampus or cortex. *p < 0.05 (Unpaired Student's *t* test). Data are presented as means ± SEMs. Each dot represents one slice (n = 7-9). CONV: conventional. SOPF: Specific and Opportunistic Pathogen-Free.

## 4. Discussion

In this study, we investigated the impact of the sanitary status of the animal facility on the composition of the faecal microbiota of 5XFAD mice, and whether the transplantation of their dysbiotic microbiota into naive mice affected cognitive performance, depending on the breeding facility of the donor mice.

The importance of gut microbiota in brain physiopathology is raising an increasing interest, especially in AD (Varesi et al., 2022). Several studies have reported intestinal dysbiosis in AD patients and transgenic mouse models of AD (Cattaneo et al., 2017; Chen et al., 2020c; Lee et al., 2018; Sun et al., 2019b; Vogt et al., 2017; Zhuang et al., 2018). Dysbiosis of the gut microbiota can result from well-described external factors, such as mode of delivery, diet, physical activity, or antibiotic use (Brown et al., 2012; Dziewiecka et al., 2022; Kesavelu and Jog, 2023; Zhang et al., 2021). Children inherit part of their mother's microbiome (Ferretti et al., 2018), and the first events that might influence the development of a dysbiotic microbiota occur before birth during *in utero* development. E. Hsiao and colleagues have shown that maternal immune activation leads to alterations in the composition of the gut microbiota in the offspring (Hsiao et al., 2013). In addition, mood disorders, such as anxiety and stress during pregnancy, result in the development of a dysbiotic microbiota in newborn mice (Galley et al., 2023). In the context of AD, there is a reciprocal relationship between the gut microbiota dysbiosis and the pathology. Modulation of the microbiota by antibiotic treatment reduces the amyloid load in the brains of treated AD mice (Guilherme et al., 2021; Minter et al., 2016) and, in a mirror image, the injection of A$\beta$ into the lateral ventricle of mice is able to induce an alteration of the gut microbiota 4 weeks later (Qian et al., 2022). Alterations of the patient lifestyle by AD, such as circadian cycle disruption, change of eating habits and physical inactivity have been shown to induce gut microbiota dysbiosis (Ikeda et al., 2002; Matenchuk et al., 2020; Xu et al., 2023). Transplantation of microbiota from AD patients into rats induces memory impairment that correlate with the clinical scores of the donors (Grabrucker et al., 2023).

In the present work, we confirmed that 5XFAD mice display age-related change in faecal bacterial composition compared to WT littermates. It should be noted that all matings were performed by pairing 5XFAD males with WT females. Therefore, all newborns were derived from naive mice, meaning that the only difference between 5XFAD mice at birth was the presence of the *APP* and *PS1* transgenes, suggesting that the production of human amyloid peptides and their downstream effects could be responsible for the development of microbial dysbiosis. Confirming this hypothesis, a recent study showed that the knock-in AD mouse models, APP$^{NL-F}$ and APP$^{NL-G-F}$, expressing mutant human APP at physiological levels, exhibit an altered gut microbiota compared to controls (Kundu et al., 2021). Our mating strategy should have avoided any confounding effects due to the maternal microbiome. However, it has recently been shown that paternal gut microbiota can also alter the offspring phenotypes (Argaw-Denboba et al., 2024). As the mice used here were housed in the respective facilities for more than 3 generations, we should have fixed the paternal effect associated with the breading facility.

As maintaining the health status of animal models is a major concern, SOPF facilities have been implemented. We observed that in such a controlled environment, 5XFAD mice developed differences in faecal bacterial composition compared to WT littermates but did not develop dysbiosis at 6 months of age, as observed in conventional housing. These results demonstrate that an environment with restricted microbial diversity can prevent the development of a known phenotype. However, breeding SOPF-born 5XFAD mice in a conventional facility restored the dysbiosis (Fig. 4). This finding is supported by data showing that co-housing conventionally raised animals with SPF animals significantly increases the diversity of the latter's gut microbiota (Chen et al., 2020b).

We then assessed the impact of faecal microbiota transplantation from 5XFAD mice on cognition in naive WT mice. Importantly, transplantation was performed without depletion of the endogenous microbiota, but repeatedly over a 2-month period to mimic a slow and chronic deregulation of the microbiota and to avoid the use of antibiotics that would interfere with the composition of the microbiota. When donor mice are bred in a conventional environment, we observed that the dysbiotic microbiota from 2- and 6-month-old 5XFAD mice induces cognitive impairment in WT mice while the microbiota of age-matched SOPF-bred 5XFAD mice did not impair memory in recipient mice. As the faecal microbiota of conventionally-bred 5XFAD mice is not very different from that of WT mice, we hypothesize that the negative effect on memory of naïve recipient may be due to differences in the faecal metabolic profile, which was not studied here (Chen et al., 2024; Choi and Mook-Jung, 2023). In SOPF-bred 5XFAD mice, the absence of a dysbiotic profile delays the phenotype of the model and might explain the lack of effect of microbiota transfer on cognitive performance. Similar experiments evaluating the cognition in WT mice receiving faecal microbiota from 5XFAD mice purchased from the Jackson Laboratory (Bar Harbor, ME, USA), a procedure reminiscent of our second set of experiments performed with SOPF-bred donors, showed that the FMT from 9 month-old mice to recipient mice depleted of the endogenous microbiota with an antibiotic cocktail, leads to deficits in spatial learning and memory in Morris water maze test in these mice (Kim et al., 2021). These experiments are difficult to compare with the present study as experimental paradigms are different and microbiota composition was assessed after or before FMT (in our case, no sample were taken post-FMT). However, a striking difference lies in the amount of *Prevotella* genus, which was found to be reduced in mice subjected to 5XFAD-FMT, compared to WT-FMT in the study of Kim *et al.* (Kim et al., 2021). In contrast, we found an increase in *Prevotella* in 5XFAD compared to WT, as described by others (Chen et al., 2020a). The housing environment of the recipient mice is also an important factor, as obesity studies have demonstrated that transplantation of human microbiota partially recapitulates obese phenotypes in conventionally reared mice, in contrast to SPF-bred mice, which show no change (Kaiser et al., 2021).

The gut microbiota composition also correlates with amyloid pathology in preclinical models and patients (Cattaneo et al., 2017; Vogt et al., 2017). Harach *et al.* showed that amyloid levels were lower when AD mice were bred in germ-free facility than in a conventional environment, and that colonization of axenic AD mice with microbiota from conventionally-bred mice was able to increase brain amyloid pathology (Harach et al., 2017). Similarly, in our study we observed that 6-month-old SOPF-bred 5XFAD mice showed a decrease in amyloid load in the hippocampus and cortex compared to conventionally bred mice of same age.

In summary, the sanitary status of the animal facility has a strong impact on AD development and associated behavioural deficits in mouse models. As the environment modifies the microbiota and its impact on diseases, different environments will drive divergent phenotypes in animal models. AD mice housed in highly controlled environments (SPF and SOPF) will have an impoverished microbiota and develop less pronounced disease phenotype than the same mice housed in conventional animal facility. The integration of these data may lead to reconsider the use of animals from conventional housing or colonized with wild microbiota (Rosshart et al., 2017) rather than SPF- and SOPF-status animals in studies investigating the relationship between gut microbiota and neurological disorders, such as AD. The translational value of preclinical studies on these complex pathologies should be enhanced by working with models that more closely mimic the natural microbiota.

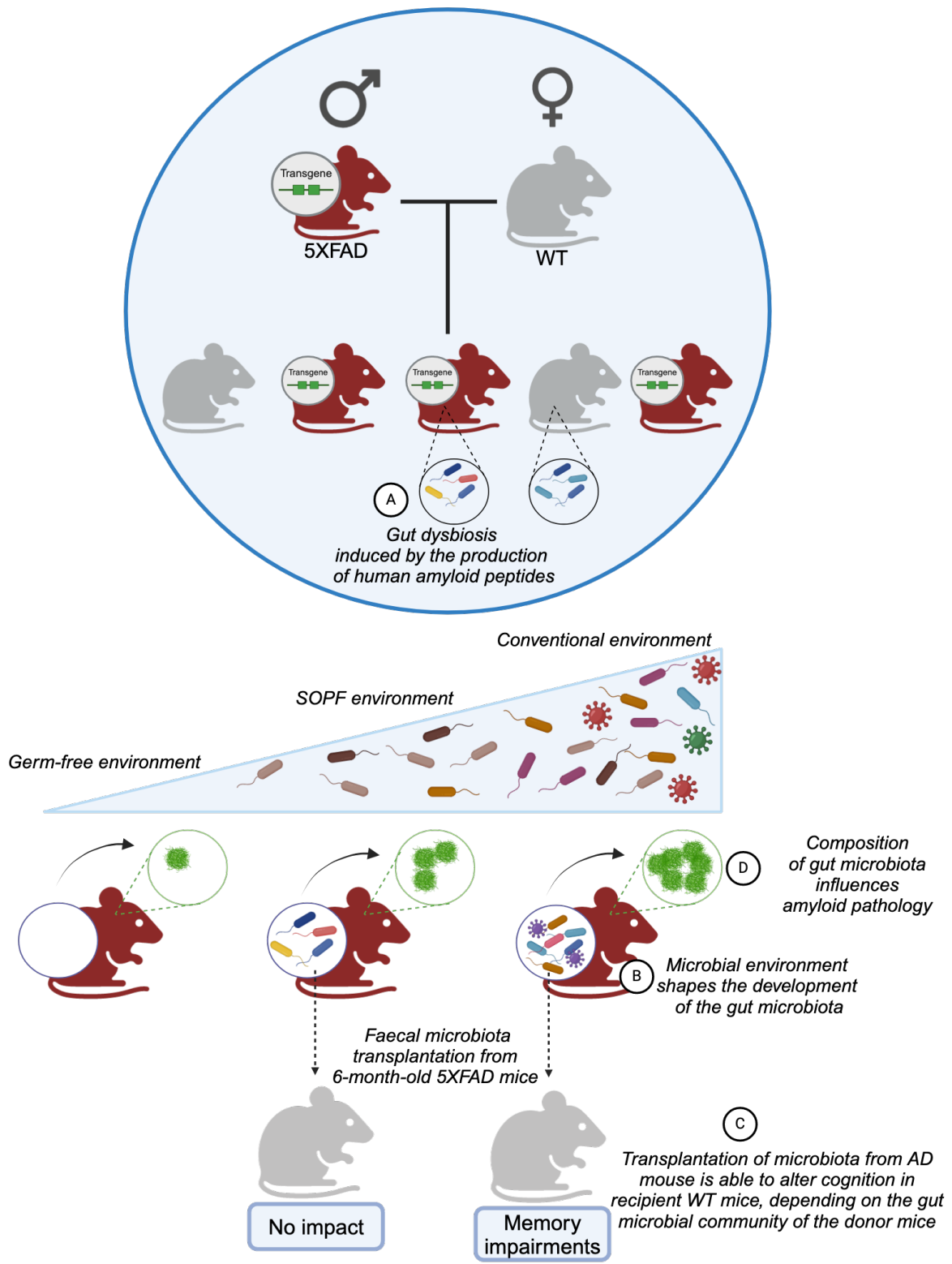

**Figure 5. Graphical abstract. (A)** The transgene in 5XFAD mice induces the development of dysbiotic faecal microbiota compared to WT littermates. **(B)** Environmental microbial diversity influences the development of gut microbiota composition. **(C)** The effect of faecal microbiota transplantation from 6-month-old 5XFAD mice on cognition in WT mice depends on the environment and the donor microbiota. **(D)** Gut microbiota has an impact on amyloid pathology in 5XFAD mice. SOPF: Specific and Opportunistic Pathogen-Free.

**Author contributions**

The authors declare no conflict of interest and approve the final version of the manuscript. CIW: Conceptualization, Investigation, Visualization, Writing - Original Draft. PG: Conceptualization, Methodology. FS: Formal analysis, Visualization. LNM: Investigation. KB: Resources. SR: Resources, Funding acquisition. PM: Supervision, Writing- Reviewing and Editing. BL: Data Curation, Writing- Reviewing and Editing. SC: Conceptualization, Supervision, Funding acquisition, Writing- Reviewing and Editing.

**Acknowledgments**

We thank Réseau des Animaleries de Montpellier, the iExplore platform and the former Animalerie of the Faculté de Médecine de Marseille for the breeding and management of mouse colonies.

This work was supported by funding from the Inserm Cross-Cutting program on Microbiota and the FEDER/Région Occitanie (call "Recherche & Société(s) 2018", MICMALZ project) to SC. This work was also supported by ANR MAD5 to SR. The PhD fellowship of C.I.W. was funded by SynAging, France Alzheimer Association and Soroptimist International (Club de Montpellier Metropole).

| Bacteria | Endo and ectoparasites | Viruses |
|---|---|---|
| *Bordetella bronchiseptica* | *Aspiculuris tetraptera* | Hantavirus |
| Cilia-associated respiratory bacillus | *Chilomastix* | K virus |
| *Citrobacter rodentium* | *Cryptosporidium muris* | Lactate dehydrogenase elevating virus |
| *Clostridium piliforme* | *Cryptosporidium parvum* | Lymphocytic choriomeningitis virus |
| *Corynebacterium bovis* | *Demodex* | MKPV |
| *Corynebacterium kutscheri* | Dermatophytes | Mouse adenovirus type 1 (FL) |
| *Helicobacter spp* | *Eimeria sp* | Mouse adenovirus type 2 (K87) |
| *Klebsiella pneumoniae/oxytoca* | *Encephalitozoon cuniculi* | Mouse cytomegalovirus |
| *Mycoplasma pulmonis* | *Entamoeba muris* | Mouse hepatitis virus |
| *Pasteurella spp* | *Giardia* muris | Mouse polyomavirus |
| *Proteus mirabilis* | *Hymenolepis diminuta* | Mouse rotavirus |
| *Pseudomonas aeruginosa* | *Hymenolepis nana* | Mouse thymic virus |
| *Salmonella spp* | *Myobia musculi* | Mousepox (*Ectromelia*) virus |
| *Staphylococcus aureus* | *Myocoptes musculinus* | Murine norovirus |
| *Streptobacillus moniliformis* | Non-pathogenic protozoa (such as *Hexamastix*, *Tritrichomonas*, etc.) | Parvovirus: Minute virus of mice Mouse parvovirus |
| Streptococci β-hemolytic (not group D) | *Pneumocystis* murina | Pneumonia virus of mice |
| *Streptococcus pneumoniae* | *Polyplax species* | Reovirus type 3 |
| | *Radfordia spp* | Sendai virus |
| | *Spironucleus* muris | Theiler's murine encephalomyelitis virus |
| | *Syphacia* sp | |
| | *Trichomonas* spp | |

**Supplementary Table 1. Exclusion list of the Plateau Central d'Elevage et d'Archivage, CNRS, Montpellier (SOPF animal facility).** Detection tests include RT-PCR, ELISA or Multiplexed Fluorometric ImmunoAssay (Bead).

## 5. References

Argaw-Denboba, A., Schmidt, T.S.B., Di Giacomo, M., Ranjan, B., Devendran, S., Mastrorilli, E., Lloyd, C.T., Pugliese, D., Paribeni, V., Dabin, J., Pisaniello, A., Espinola, S., Crevenna, A., Ghosh, S., Humphreys, N., Boruc, O., Sarkies, P., Zimmermann, M., Bork, P., Hackett, J.A., 2024. Paternal microbiome perturbations impact offspring fitness. Nature 629, 652-659.

Bevins, R.A., Besheer, J., 2006. Object recognition in rats and mice: a one-trial non-matching-to-sample learning task to study 'recognition memory'. Nat Protoc 1, 1306-1311.

Boehme, M., Guzzetta, K.E., Bastiaanssen, T.F.S., van de Wouw, M., Moloney, G.M., Gual-Grau, A., Spichak, S., Olavarria-Ramirez, L., Fitzgerald, P., Morillas, E., Ritz, N.L., Jaggar, M., Cowan, C.S.M., Crispie, F., Donoso, F., Halitzki, E., Neto, M.C., Sichetti, M., Golubeva, A.V., Fitzgerald, R.S., Claesson, M.J., Cotter, P.D., O'Leary, O.F., Dinan, T.G., Cryan, J.F., 2021. Microbiota from young mice counteracts selective age-associated behavioral deficits. Nat Aging 1, 666-676.

Boulanger, N., Insonere, J.L., Van Blerk, S., Barthel, C., Serres, C., Rais, O., Roulet, A., Servant, F., Duron, O., Lelouvier, B., 2023. Cross-alteration of murine skin and tick microbiome concomitant with pathogen transmission after Ixodes ricinus bite. Microbiome 11, 250.

Brown, K., DeCoffe, D., Molcan, E., Gibson, D.L., 2012. Diet-induced dysbiosis of the intestinal microbiota and the effects on immunity and disease. Nutrients 4, 1095-1119.

Cattaneo, A., Cattane, N., Galluzzi, S., Provasi, S., Lopizzo, N., Festari, C., Ferrari, C., Guerra, U.P., Paghera, B., Muscio, C., Bianchetti, A., Volta, G.D., Turla, M., Cotelli, M.S., Gennuso, M., Prelle, A., Zanetti, O., Lussignoli, G., Mirabile, D., Bellandi, D., Gentile, S., Belotti, G., Villani, D., Harach, T., Bolmont, T., Padovani, A., Boccardi, M., Frisoni, G.B., Group, I.-F., 2017. Association of brain amyloidosis with pro-inflammatory gut bacterial taxa and peripheral inflammation markers in cognitively impaired elderly. Neurobiol Aging 49, 60-68.

Chen, C., Ahn, E.H., Kang, S.S., Liu, X., Alam, A., Ye, K., 2020a. Gut dysbiosis contributes to amyloid pathology, associated with C/EBPbeta/AEP signaling activation in Alzheimer's disease mouse model. Sci Adv 6, eaba0466.

Chen, J., Zhang, S., Feng, X., Wu, Z., Dubois, W., Thovarai, V., Ahluwalia, S., Gao, S., Chen, J., Peat, T., Sen, S.K., Trinchieri, G., Young, N.S., Mock, B.A., 2020b. Conventional Co-Housing Modulates Murine Gut Microbiota and Hematopoietic Gene Expression. Int J Mol Sci 21.

Chen, Y., Fang, L., Chen, S., Zhou, H., Fan, Y., Lin, L., Li, J., Xu, J., Chen, Y., Ma, Y., Chen, Y., 2020c. Gut Microbiome Alterations Precede Cerebral Amyloidosis and Microglial Pathology in a Mouse Model of Alzheimer's Disease. Biomed Res Int 2020, 8456596.

Chen, Y., Li, Y., Fan, Y., Chen, S., Chen, L., Chen, Y., Chen, Y., 2024. Gut microbiota-driven metabolic alterations reveal gut-brain communication in Alzheimer's disease model mice. Gut Microbes 16, 2302310.

Choi, H., Mook-Jung, I., 2023. Functional effects of gut microbiota-derived metabolites in Alzheimer's disease. Curr Opin Neurobiol 81, 102730.

Dodiya, H.B., Kuntz, T., Shaik, S.M., Baufeld, C., Leibowitz, J., Zhang, X., Gottel, N., Zhang, X., Butovsky, O., Gilbert, J.A., Sisodia, S.S., 2019. Sex-specific effects of microbiome perturbations on cerebral Abeta amyloidosis and microglia phenotypes. J Exp Med 216, 1542-1560.

Dziewiecka, H., Buttar, H.S., Kasperska, A., Ostapiuk-Karolczuk, J., Domagalska, M., Cichon, J., Skarpanska-Stejnborn, A., 2022. Physical activity induced alterations of gut microbiota in humans: a systematic review. BMC Sports Sci Med Rehabil 14, 122.

Escudie, F., Auer, L., Bernard, M., Mariadassou, M., Cauquil, L., Vidal, K., Maman, S., Hernandez-Raquet, G., Combes, S., Pascal, G., 2018. FROGS: Find, Rapidly, OTUs with Galaxy Solution. Bioinformatics 34, 1287-1294.

Ferretti, P., Pasolli, E., Tett, A., Asnicar, F., Gorfer, V., Fedi, S., Armanini, F., Truong, D.T., Manara, S., Zolfo, M., Beghini, F., Bertorelli, R., De Sanctis, V., Bariletti, I., Canto, R., Clementi, R., Cologna, M., Crifo, T., Cusumano, G., Gottardi, S., Innamorati, C., Mase, C., Postai, D., Savoi, D., Duranti, S., Lugli, G.A., Mancabelli, L., Turroni, F., Ferrario, C., Milani, C., Mangifesta, M., Anzalone, R., Viappiani, A., Yassour, M., Vlamakis, H., Xavier, R., Collado, C.M., Koren, O., Tateo, S., Soffiati, M., Pedrotti, A., Ventura, M., Huttenhower, C., Bork, P., Segata, N., 2018. Mother-to-Infant Microbial Transmission from Different Body Sites Shapes the Developing Infant Gut Microbiome. Cell Host Microbe 24, 133-145 e135.

Galley, J.D., Mashburn-Warren, L., Blalock, L.C., Lauber, C.L., Carroll, J.E., Ross, K.M., Hobel, C., Coussons-Read, M., Dunkel Schetter, C., Gur, T.L., 2023. Maternal anxiety, depression and stress affects offspring gut microbiome diversity and bifidobacterial abundances. Brain Behav Immun 107, 253-264.

Grabrucker, S., Marizzoni, M., Silajdzic, E., Lopizzo, N., Mombelli, E., Nicolas, S., Dohm-Hansen, S., Scassellati, C., Moretti, D.V., Rosa, M., Hoffmann, K., Cryan, J.F., O'Leary, O.F., English, J.A., Lavelle, A., O'Neill, C., Thuret, S., Cattaneo, A., Nolan, Y.M., 2023. Microbiota from Alzheimer's patients induce deficits in cognition and hippocampal neurogenesis. Brain 146, 4916-4934.

Guilherme, M.D.S., Nguyen, V.T.T., Reinhardt, C., Endres, K., 2021. Impact of Gut Microbiome Manipulation in 5xFAD Mice on Alzheimer's Disease-Like Pathology. Microorganisms 9.

Guo, M., Peng, J., Huang, X., Xiao, L., Huang, F., Zuo, Z., 2021. Gut Microbiome Features of Chinese Patients Newly Diagnosed with Alzheimer's Disease or Mild Cognitive Impairment. J Alzheimers Dis 80, 299-310.

Harach, T., Marungruang, N., Duthilleul, N., Cheatham, V., Mc Coy, K.D., Frisoni, G., Neher, J.J., Fak, F., Jucker, M., Lasser, T., Bolmont, T., 2017. Reduction of Abeta amyloid pathology in APPPS1 transgenic mice in the absence of gut microbiota. Sci Rep 7, 41802.

Haran, J.P., Bhattarai, S.K., Foley, S.E., Dutta, P., Ward, D.V., Bucci, V., McCormick, B.A., 2019. Alzheimer's Disease Microbiome Is Associated with Dysregulation of the Anti-Inflammatory P-Glycoprotein Pathway. mBio 10.

Hazan, S., 2020. Rapid improvement in Alzheimer's disease symptoms following fecal microbiota transplantation: a case report. J Int Med Res 48, 300060520925930.

He, C., Jiang, J., Liu, J., Zhou, L., Ge, Y., Yang, Z., 2024. Pseudostellaria heterophylla polysaccharide mitigates Alzheimer's-like pathology via regulating the microbiota-gut-brain axis in 5 x FAD mice. Int J Biol Macromol 270, 132372.

Holsinger, R.M.D., Elangovan, S., 2020. Neuroprotective effects of fecal microbiota transplantation in a mouse model of Alzheimer's disease. Alzheimer's & Dementia 16, e046523.

Hsiao, E.Y., McBride, S.W., Hsien, S., Sharon, G., Hyde, E.R., McCue, T., Codelli, J.A., Chow, J., Reisman, S.E., Petrosino, J.F., Patterson, P.H., Mazmanian, S.K., 2013. Microbiota modulate behavioral and physiological abnormalities associated with neurodevelopmental disorders. Cell 155, 1451-1463.

Ikeda, M., Brown, J., Holland, A.J., Fukuhara, R., Hodges, J.R., 2002. Changes in appetite, food preference, and eating habits in frontotemporal dementia and Alzheimer's disease. J Neurol Neurosurg Psychiatry 73, 371-376.

Ismeurt, C., Giannoni, P., Claeysen, S., 2020. Chapter 13 - The 5×FAD mouse model of Alzheimer's disease. In: Martin, C.R., Preedy, V.R. (Eds.), Diagnosis and Management in Dementia. Academic Press, pp. 207-221.

Kaiser, T., Nalluri, H., Zhu, Z., Staley, C., 2021. Donor Microbiota Composition and Housing Affect Recapitulation of Obese Phenotypes in a Human Microbiota-Associated Murine Model. Front Cell Infect Microbiol 11, 614218.

Kesavelu, D., Jog, P., 2023. Current understanding of antibiotic-associated dysbiosis and approaches for its management. Ther Adv Infect Dis 10, 20499361231154443.

Kim, N., Jeon, S.H., Ju, I.G., Gee, M.S., Do, J., Oh, M.S., Lee, J.K., 2021. Transplantation of gut microbiota derived from Alzheimer's disease mouse model impairs memory function and neurogenesis in C57BL/6 mice. Brain Behav Immun 98, 357-365.

Kobayashi, Y., Sugahara, H., Shimada, K., Mitsuyama, E., Kuhara, T., Yasuoka, A., Kondo, T., Abe, K., Xiao, J.Z., 2017. Therapeutic potential of Bifidobacterium breve strain A1 for preventing cognitive impairment in Alzheimer's disease. Sci Rep 7, 13510.

Kundu, P., Torres, E.R.S., Stagaman, K., Kasschau, K., Okhovat, M., Holden, S., Ward, S., Nevonen, K.A., Davis, B.A., Saito, T., Saido, T.C., Carbone, L., Sharpton, T.J., Raber, J., 2021. Integrated analysis of behavioral, epigenetic, and gut microbiome analyses in App(NL-G-F), App(NL-F), and wild type mice. Sci Rep 11, 4678.

Lee, H.J., Hwang, Y.H., Kim, D.H., 2018. Lactobacillus plantarum C29-Fermented Soybean (DW2009) Alleviates Memory Impairment in 5XFAD Transgenic Mice by Regulating Microglia Activation and Gut Microbiota Composition. Mol Nutr Food Res 62, e1800359.

Li, B., He, Y., Ma, J., Huang, P., Du, J., Cao, L., Wang, Y., Xiao, Q., Tang, H., Chen, S., 2019. Mild cognitive impairment has similar alterations as Alzheimer's disease in gut microbiota. Alzheimers Dement 15, 1357-1366.

Ling, Z., Zhu, M., Yan, X., Cheng, Y., Shao, L., Liu, X., Jiang, R., Wu, S., 2020. Structural and Functional Dysbiosis of Fecal Microbiota in Chinese Patients With Alzheimer's Disease. Front Cell Dev Biol 8, 634069.

Liu, P., Wu, L., Peng, G., Han, Y., Tang, R., Ge, J., Zhang, L., Jia, L., Yue, S., Zhou, K., Li, L., Luo, B., Wang, B., 2019. Altered microbiomes distinguish Alzheimer's disease from amnestic mild cognitive impairment and health in a Chinese cohort. Brain Behav Immun 80, 633-643.

Lluch, J., Servant, F., Paisse, S., Valle, C., Valiere, S., Kuchly, C., Vilchez, G., Donnadieu, C., Courtney, M., Burcelin, R., Amar, J., Bouchez, O., Lelouvier, B., 2015. The Characterization of Novel Tissue Microbiota Using an Optimized 16S Metagenomic Sequencing Pipeline. PLoS One 10, e0142334.

Luo, S., Zhang, X., Huang, S., Feng, X., Zhang, X., Xiang, D., 2022. A monomeric polysaccharide from Polygonatum sibiricum improves cognitive functions in a model of Alzheimer's disease by reshaping the gut microbiota. Int J Biol Macromol 213, 404-415.

Matenchuk, B.A., Mandhane, P.J., Kozyrskyj, A.L., 2020. Sleep, circadian rhythm, and gut microbiota. Sleep Med Rev 53, 101340.

Minter, M.R., Hinterleitner, R., Meisel, M., Zhang, C., Leone, V., Zhang, X., Oyler-Castrillo, P., Zhang, X., Musch, M.W., Shen, X., Jabri, B., Chang, E.B., Tanzi, R.E., Sisodia, S.S., 2017. Antibiotic-induced perturbations in microbial diversity during post-natal development alters amyloid pathology in an aged APP(SWE)/PS1(DeltaE9) murine model of Alzheimer's disease. Sci Rep 7, 10411.

Minter, M.R., Zhang, C., Leone, V., Ringus, D.L., Zhang, X., Oyler-Castrillo, P., Musch, M.W., Liao, F., Ward, J.F., Holtzman, D.M., Chang, E.B., Tanzi, R.E., Sisodia, S.S., 2016. Antibiotic-induced perturbations in gut microbial diversity influences neuro-inflammation and amyloidosis in a murine model of Alzheimer's disease. Sci Rep 6, 30028.

Oakley, H., Cole, S.L., Logan, S., Maus, E., Shao, P., Craft, J., Guillozet-Bongaarts, A., Ohno, M., Disterhoft, J., Van Eldik, L., Berry, R., Vassar, R., 2006. Intraneuronal beta-amyloid aggregates, neurodegeneration, and neuron loss in transgenic mice with five familial Alzheimer's disease mutations: potential factors in amyloid plaque formation. J Neurosci 26, 10129-10140.

Park, S.H., Lee, J.H., Shin, J., Kim, J.S., Cha, B., Lee, S., Kwon, K.S., Shin, Y.W., Choi, S.H., 2021. Cognitive function improvement after fecal microbiota transplantation in Alzheimer's dementia patient: a case report. Curr Med Res Opin 37, 1739-1744.

Qian, X.H., Liu, X.L., Chen, G., Chen, S.D., Tang, H.D., 2022. Injection of amyloid-beta to lateral ventricle induces gut microbiota dysbiosis in association with inhibition of cholinergic anti-inflammatory pathways in Alzheimer's disease. J Neuroinflammation 19, 236.

Rezaei Asl, Z., Sepehri, G., Salami, M., 2019. Probiotic treatment improves the impaired spatial cognitive performance and restores synaptic plasticity in an animal model of Alzheimer's disease. Behav Brain Res 376, 112183.

Rosshart, S.P., Vassallo, B.G., Angeletti, D., Hutchinson, D.S., Morgan, A.P., Takeda, K., Hickman, H.D., McCulloch, J.A., Badger, J.H., Ajami, N.J., Trinchieri, G., Pardo-Manuel de Villena, F., Yewdell, J.W., Rehermann, B., 2017. Wild Mouse Gut Microbiota Promotes Host Fitness and Improves Disease Resistance. Cell 171, 1015-1028 e1013.

Sun, B.L., Li, W.W., Wang, J., Xu, Y.L., Sun, H.L., Tian, D.Y., Wang, Y.J., Yao, X.Q., 2019a. Gut Microbiota Alteration and Its Time Course in a Tauopathy Mouse Model. J Alzheimers Dis 70, 399-412.

Sun, J., Xu, J., Ling, Y., Wang, F., Gong, T., Yang, C., Ye, S., Ye, K., Wei, D., Song, Z., Chen, D., Liu, J., 2019b. Fecal microbiota transplantation alleviated Alzheimer's disease-like pathogenesis in APP/PS1 transgenic mice. Transl Psychiatry 9, 189.

Varesi, A., Pierella, E., Romeo, M., Piccini, G.B., Alfano, C., Bjorklund, G., Oppong, A., Ricevuti, G., Esposito, C., Chirumbolo, S., Pascale, A., 2022. The Potential Role of Gut Microbiota in Alzheimer's Disease: From Diagnosis to Treatment. Nutrients 14.

Vogt, N.M., Kerby, R.L., Dill-McFarland, K.A., Harding, S.J., Merluzzi, A.P., Johnson, S.C., Carlsson, C.M., Asthana, S., Zetterberg, H., Blennow, K., Bendlin, B.B., Rey, F.E., 2017. Gut microbiome alterations in Alzheimer's disease. Sci Rep 7, 13537.

Xu, L., Li, W., Ling, L., Zhang, Z., Cui, Z., Ge, J., Wang, Y., Meng, Q., Wang, Y., Liu, K., Zhou, J., Zeng, F., Wang, J., Wu, J., 2023. A Sedentary Lifestyle Changes the Composition and Predicted Functions of the Gut Bacterial and Fungal Microbiota of Subjects from the Same Company. Curr Microbiol 80, 368.

Zhang, C., Li, L., Jin, B., Xu, X., Zuo, X., Li, Y., Li, Z., 2021. The Effects of Delivery Mode on the Gut Microbiota and Health: State of Art. Front Microbiol 12, 724449.

Zhuang, Z.Q., Shen, L.L., Li, W.W., Fu, X., Zeng, F., Gui, L., Lu, Y., Cai, M., Zhu, C., Tan, Y.L., Zheng, P., Li, H.Y., Zhu, J., Zhou, H.D., Bu, X.L., Wang, Y.J., 2018. Gut Microbiota is Altered in Patients with Alzheimer's Disease. J Alzheimers Dis 63, 1337-1346.